\documentclass[8pt,a4paper,twocolumn]{article}
\usepackage[margin=20mm]{geometry}
\usepackage{cite}
\usepackage{amsmath,amssymb,amsfonts}
\usepackage[ruled]{algorithm2e} % For algorithms
\usepackage{graphicx}
\usepackage{textcomp}
\usepackage{xcolor}
\def\BibTeX{{\rm B\kern-.05em{\sc i\kern-.025em b}\kern-.08em
　　T\kern-.1667em\lower.7ex\hbox{E}\kern-.125emX}}
\begin{document}

\title{\Huge Real-time Trading System based on Selections of Potentially Profitable, Uncorrelated, and Balanced Stocks by NP-hard Combinatorial Optimization}

\author{
Kosuke Tatsumura$^{\ast}$, Ryo Hidaka, Jun Nakayama, \\
Tomoya Kashimata, and Masaya Yamasaki \\
\small Corporate Research and Development Center, Toshiba Corporation, Japan\\
$^{\ast}$\small Corresponding author: Kosuke Tatsumura (e-mail: kosuke.tatsumura@toshiba.co.jp)
}
\date{}

\maketitle

\begin{abstract}
Financial portfolio construction problems are often formulated as quadratic and discrete (combinatorial) optimization that belong to the nondeterministic polynomial time (NP)-hard class in computational complexity theory. Ising machines are hardware devices that work in quantum-mechanical/quantum-inspired principles for quickly solving NP-hard optimization problems, which potentially enable making trading decisions based on NP-hard optimization in the time constraints for high-speed trading strategies. Here we report a real-time stock trading system that determines long(buying)/short(selling) positions through NP-hard portfolio optimization for improving the Sharpe ratio using an embedded Ising machine based on a quantum-inspired algorithm called simulated bifurcation. The Ising machine selects a balanced (delta-neutral) group of stocks from an $N$-stock universe according to an objective function involving maximizing instantaneous expected returns defined as deviations from volume-weighted average prices and minimizing the summation of statistical correlation factors (for diversification). It has been demonstrated in the Tokyo Stock Exchange that the trading strategy based on NP-hard portfolio optimization for $N$=128 is executable with the FPGA (field-programmable gate array)-based trading system with a response latency of 164 $\mu$s.
\end{abstract}

\section{Introduction}\label{sec:introduction}
Many portfolio construction/selection problems in finance are, with considering minimum transaction lots or other discretenesses of decision variables as realistic constraints, known to be nondeterministic polynomial (NP)-hard in computer science~\cite{bienstock96, mansini99}. Those include discrete optimizations of Markowitz's mean-variance model~\cite{markowitz52} for better risk-return characteristics~\cite{venturelli19,lang22}, multi-period portfolio optimizations (or optimal trading trajectory problems)~\cite{rosenberg15,steinhauer20,mugel22}, and correlation-diversified portfolio constructions including maximum independent set (MIS) problem-based ones~\cite{butenko03,boginski04,marzec16} and permutation of assets-based one~\cite{sakurai21}.

Recently, special-purpose computers for NP-hard combinatorial (or discrete) optimization, called Ising machines~\cite{sbm1, FPL19, sbm2, NatEle, kanao23, johnson11,king23, honjo21, pierangeli19, cai20, aadit22, moy22, sharma22, takemoto19, kawamura23, matsubara20, waidyasooriya21, okuyama19}, have attracted intense attention. Ising problems are the ground (energy minimum)-state search problems of Ising spin models, which consist of binary variables, called spins, coupled each other with pairwise interactions. The Ising problem belongs to the NP-hard class~\cite{barahona82, lucas14}; a variety of notoriously hard problems can be represented in the form of the Ising problem~\cite{lucas14}. The Ising machine is a heuristic methodology and searches for the optimal (exact) or near-optimal solutions of the Ising problem in the whole solution space. Many Ising machines have claimed higher speed performance than simulated annealing~\cite{SA83} (on von Neumann computers), a conventional heuristic for combinatorial optimization.

The Ising machines are implemented with various hardware including superconducting flux qubits~\cite{johnson11,king23}, hybrid electronic-optical systems~\cite{honjo21, pierangeli19}, memristor-based neural networks~\cite{cai20}, probabilistic bits~\cite{aadit22}, coupled-oscillator circuits~\cite{moy22}, analog computing units~\cite{sharma22}, application specific integrated circuits (ASICs)~\cite{takemoto19,kawamura23,matsubara20}, field programmable gate array (FPGAs)~\cite{sbm1, FPL19, sbm2, NatEle, waidyasooriya21}, and graphics processing units (GPUs)~\cite{sbm1, sbm2, okuyama19}.

The Ising machines may enable making more rational judgments based on NP-hard combinatorial optimizations for automated trading systems~\cite{yoo23,fil20,huang19,denholm15,leber11} that become increasingly important in financial markets~\cite{malceniece23,brogaard14}. Those trading systems are typical real-time systems that must respond (sense, judge, and react) within critically defined time constraints. Many high-speed trading systems~\cite{yoo23,fil20,huang19,denholm15,leber11} utilize FPGAs to shorten the latency from the market feed arrival to order packet issuance. Thus, among various Ising machines, FPGA-based ones (Ising machines that can be accelerated with modern FPGA architectures~\cite{betz12}) are suitable for high-speed trading systems because they can be embedded together with other system components in the FPGAs. The trading systems utilizing FPGA-based embeddable Ising machines as in~\cite{ISCAS20} have been, however, not extensively studied. Furthermore, the execution capability of such a trading system needs to be validated in the actual market because of the latency of the system and the lifetime of the trading opportunity depending on the activities of other trading entities.

Here we propose a trading strategy based on selections of potentially profitable, uncorrelated, and balanced stocks by NP-hard combinatorial optimization and show through real-time trading that the strategy is executable with an automated real-time system using an FPGA-based embedded Ising machine for the discrete selection problem.

Based on the demand in the direction of convergence of the stock price to the volume-weighted average price (VWAP), the proposed strategy considers the deviations of stock prices from the VWAPs as instantaneous expected returns and selects a balanced (delta-neutral) group of stocks from an $N$-stock universe according to an objective function involving maximizing the expected returns and minimizing the summation of statistical correlation factors (for correlation-diversification). The selection problem is formulated as quadratic and discrete optimization and solved by an Ising machine based on a quantum-inspired algorithm called simulated bifurcation (SB)~\cite{sbm1, FPL19, sbm2, NatEle, kanao23}. SB was derived from a classical counterpart to a quantum adiabatic optimization method called a quantum bifurcation machine~\cite{qbm} and numerically simulates the adiabatic time-evolution of a classical nonlinear oscillator network exhibiting bifurcation phenomena, where two branches of the bifurcation in each oscillator correspond to two states of each Ising spin. To reduce the system-wide latency by decreasing the input data size of the SB machine (SBM, a hardware implementation of SB) from $\mathcal{O}(N^2)$ to $\mathcal{O}(N)$, we separate the data describing the problem into two components that change tick-by-tick or day-by-day and customize the basic SBM design~\cite{FPL19}. We discuss the execution capability of the system by comparing the real-time transaction records of the system in the Tokyo Stock Exchange (TSE) with a backcast simulation of the strategy assuming the orders issued are necessarily filled.

The rest of the paper is organized as follows. In Sec.~\ref{sec_strategy} (trading strategy), we describe the proposed strategy and formulate the discrete selection problem in the forms of quadratic unconstrained binary optimization (QUBO) and the Ising problem. Sec.~\ref{Sec_System} (system) describes the architecture of the system, the customization of the SBM core, and the implementation details. Sec.~\ref{sec_expt} (experiment) describes the transaction records in the TSE and the execution capability of the system.

\section{Trading strategy}\label{sec_strategy}
\subsection{Discrete optimization-based strategy}
The proposed strategy considers the deviations of stock prices from the VWAPs as instantaneous expected returns and bets that the deviations would eventually converge (partially) in the trading hours. To improve the reward-to-variability ratio (or the Sharpe ratio~\cite{sharpe66, backus93}), it simultaneously holds multiple positions selected through a discrete portfolio optimization problem making the group of positions being market-neutral~\cite{gatev06} and correlation-diversified~\cite{butenko03,boginski04,marzec16,sakurai21}. 

There is demand in the direction of convergence of the stock price to the VWAP~\cite{berkowitz88,kakade04,bialkowski08}. For institutional investors mainly through passive investments, one of the common methods for reducing the trading impact on market prices is that the fund managers, with a certain fee promised, ask brokerages to execute their large volume trades on the VWAP determined at the end of the trading hours. If the average executed price is the same as the end-of-trading-hours VWAP, the brokerage earns the fee. If the brokerage executes the trades at prices more favorable than the VWAP, this brokerage earns more than the fee.

Considering the deviations of stock prices from the VWAP as expected returns, the strategy takes long positions of the underperforming stocks and short positions of the outperforming stocks and statistically expects that the underperforming stocks would move up while the outperforming stocks would move down. To adapt to various market conditions (uptrend, downtrend, or sideways), the strategy matches long(/short) positions with short(/long) positions so that the overall deltas of the positions total almost zero (delta neutral)~\cite{gatev06}. In addition, to statistically reduce the deviation of the returns (risk), the strategy incorporates the concept of correlation-diversified portfolio~\cite{butenko03,boginski04,marzec16,sakurai21}; the multiple long/short positions are selected so that the stocks involved are uncorrelated with each other.

The Sharpe ratio~\cite{sharpe66} is, in this work, the ratio of the mean to the standard deviation of the return (the profit and loss per period for an investment) from a strategy as in~\cite{backus93}. To enhance the Sharpe ratio of the proposed strategy, a group including $N_{\mathrm{s}}$ stocks is selected from an $N$-stock universe as the candidates of open positions (positions to be taken) so that (i) the summation of instantaneous expected returns is maximized (for maximizing returns), (ii) the summation of statistical correlation factors is minimized (for diversification), and (iii) the numbers of long/short positions are equal (delta-neutral). This is a discrete optimization problem. The selection of $N_{\mathrm{s}}$-stock group is executed every time the market situation changes and then the selected group is evaluated for determining the opening.

The deviation of the stock price from the VWAP ($\Delta p_{i}$) normalized with the base price on the day ($p_{i}^{b}$) is expressed by $\Delta p_{i}=(p_{i}-VWAP_{i})/p_{i}^{b}$, where $p_{i}$ is the middle price between the best ask ($ask$) and the best bid ($bid$). When the sign of $\Delta p_{i}$ is negative (/positive), $i$th stock is the candidate of long (/short) position. The absolute value of $\Delta p_{i}$ corresponds to the instantaneous expected return of the $i$th-stock position.

The number of lots per order for a stock ($L_{i}$) is determined to make the amount of transaction ($A_\mathrm{trans}$) common for all tradable stocks by rounding with considering the minimum tradable shares per order (a lot) of the stock ($S_{i}^\mathrm{min}$) and the base price on the day ($p_{i}^{b}$); $L_{i}=\lfloor A_\mathrm{trans}/S_{i}^\mathrm{min} p_{i}^{b} \rfloor$. The number of intraday positions is controlled to be within a maximum number ($P_\mathrm{max}$) and all positions are closed (unwind) before the close of the day. Duplicate pair positions are not allowed.

In this work, the correlation factor between $i$th and $j$th stocks for a business day ($\hat{\sigma}_{i,j}$) is defined based on the price deviation sequences against the VWAP as follows.

\begin{equation}\label{Eq_sigma_day}
\hat{\sigma}_{i,j}=\frac{\displaystyle
	\sum_{k}\left(p_{i}^{k}-VWAP_{i}^{k}\right)\left(p_{j}^{k}-VWAP_{j}^{k}\right)
	}{\displaystyle
	\sum_{k}\left|p_{i}^{k}-VWAP_{i}^{k}\right|\sum_{k}\left|p_{j}^{k}-VWAP_{j}^{k}\right|},
\end{equation}
where $p_{i}^{k}$ and $VWAP_{i}^{k}$ are the middle price and VWAP of $i$th stock sampled at one-second intervals. The correlation factor ($\sigma_{i,j}$) in the strategy is the average value for the last five business days of $\hat{\sigma}_{i,j}$ and is normalized to be in $[0,1]$.

\subsection{Formulation}
The problem to select $N_{\mathrm{s}}$ stocks from an $N$-stock universe according to a cost (objective) function involving maximizing instantaneous expected returns ($\left| \Delta p_{i} \right |$) and minimizing the summation of statistical correlation factors ($\sigma_{i,j}$) of the stocks involved under the constrain for delta-neutral positions is formulated in the form of quadratic unconstrained binary optimization (QUBO).

In this subsection, we explain the QUBO formulation for explanatory clarity, but the Ising machine takes the input data for the Ising problem in a one-to-one relationship with the QUBO problem as described in the next subsection. The primitive data defining the problem ($\{\Delta p_{i}\}$ vector and $\sigma$ matrix) are converted directly to the Ising formulation in the system (not via the QUBO formulation).

Define a decision (bit) variable $b_{i}$ ($b_{i}\in{\{0,1\}}$) as taking value 1 if $i$th stock is selected and 0 otherwise. When $i$th stock is selected, the sign of $\Delta p_{i}$ [$sgn(\Delta p_{i})$] indicates whether it corresponds to a long or short position. We prepare $N$ bit variables for an $N$-stock universe.

In the QUBO formulation, we search for the bit configuration $\{b_{i}\}$ that minimizes the QUBO cost function $H_{\mathrm{QUBO}}$. $H_{\mathrm{QUBO}}$ is a linear combination of a cost function $H_{\mathrm{cost}}$ and a penalty function $H_{\mathrm{penalty}}$.

\begin{equation}\label{Eq_Hqubo}
H_{\mathrm{QUBO}}=\sum_{i}^{N}\sum_{j}^{N}Q_{i,j}b_{i}b_{j}=H_{\mathrm{cost}}+H_{\mathrm{penalty}}.
\end{equation}
The cost function to be minimized is defined by
\begin{equation}\label{Eq_qubo_cost}
H_{\mathrm{cost}}=\sum_{i}\sum_{j}Q_{i,j}^{\mathrm{cost}}b_{i}b_{j},
\end{equation}
\begin{equation}\label{Eq_Q_cost}
Q_{i,j}^{\mathrm{cost}}=
\begin{cases}
-c_{1} \left| \Delta p_{i} \right |& (\text{if}\;i=j),\\
\sigma_{i,j} & \text{(otherwise)},
\end{cases}
\end{equation}
where $c_{1}$ is a positive coefficient. Note that $b_{i}^2=b_{i}$ for diagonal terms ($i=j$). The constraints for $N_{\mathrm{s}}$-stock selection and delta-neutral positions are represented as a penalty function expressed by
\begin{equation}\label{Eq_qubo_penalty}
H_{\mathrm{penalty}}=c_{2} \left( \left(\sum_{i} b_{i}\right)-N_{s} \right)^2 +c_{3}\left(\sum_{i}sgn(\Delta p_{i})b_{i}\right)^2.
\end{equation}
where $c_{2}$ and $c_{3}$ are positive coefficients. The first and second terms correspond to the constraints for $N_{\mathrm{s}}$-stock selection and delta-neutral positions, respectively. Constraint violations increase the penalty, with $H_{\mathrm{penalty}}=0$ if there are no violations. Note that the nondiagonal elements in the coupling coefficient matrix $Q$ in Eq. \eqref{Eq_Hqubo} include not only $\sigma_{i,j}$ in Eq. \eqref{Eq_Q_cost} but also components of $sgn(\Delta p_{i})$ coming from the second term in Eq. \eqref{Eq_qubo_penalty}. QUBOs are known to be NP-hard problems for classical computers~\cite{lucas14}. Since the cost function in Eq. \eqref{Eq_Q_cost} is quadratic, the discrete optimization involved in the strategy is thought to be NP-hard problems.

\subsection{Separation of problem components}
The discrete optimization problem to be solved at a market situation is described as an $N \times N$ size of coupling coefficient matrix $Q$ (in Eq. \eqref{Eq_Hqubo}), which should be transferred to the Ising machine (in this work, SBM) every time the market situation changes. To reduce the system-wide latency by decreasing the size of data transferred from $\mathcal{O}(N^2)$ to $\mathcal{O}(N)$, we separate the data describing the problem into two components that change tick-by-tick or day-by-day. We prepare additional circuit units for the computation depending only on the tick-by-tick data to the basic SBM design (see Sec.~\ref{Sec_System}).

In the QUBO formulation in Eqs. \eqref{Eq_Hqubo}, \eqref{Eq_qubo_cost}, \eqref{Eq_Q_cost} and \eqref{Eq_qubo_penalty}, the $\{\Delta p_{i}\}$ vector is the tick-by-tick change component and the $\sigma$ matrix is the day-by-day change component. The QUBO problem can be represented in the form of the Ising problem (see APPENDIX A), where the decision variables are binary variable called spins $s_{i}$ ($s_{i}\in{\{-1,1\}}$) and the problem is represented by a coupling coefficient matrix $J$ and a bias vector $h$. We describe the Ising cost function $H_{\mathrm{Ising}}$ as a linear combination of terms that include the day-by-day change components ($J^{\mathrm{day}}$, $h^{\mathrm{day}}$) or tick-by-tick change components ($J^{\mathrm{tick}}$, $h^{\mathrm{tick}}$) as follows;
\begin{multline}
H_{\mathrm{Ising}}=
-\frac{1}{2}\sum_{i}^{N}\sum_{j}^{N}J_{i,j}^{\mathrm{day}}s_{i}s_{j}-\frac{1}{2}\sum_{i}^{N}\sum_{j}^{N}J_{i,j}^{\mathrm{tick}}s_{i}s_{j} \\
+\sum_{i}^{N}h_{i}^{\mathrm{day}}s_{i}+\sum_{i}^{N}h_{i}^{\mathrm{tick}}s_{i}.
\end{multline}
Here, the ($J^{\mathrm{day}}$, $h^{\mathrm{day}}$) and ($J^{\mathrm{tick}}$, $h^{\mathrm{tick}}$) can be calculated from the $\sigma$ matrix and $\{\Delta p_{i}\}$ vector, respectively.

The SBM core described in the next section stores the ($J^{\mathrm{day}}$, $h^{\mathrm{day}}$) and ($J^{\mathrm{tick}}$, $h^{\mathrm{tick}}$) data in the separated memories. Note that the size of $J_{i,j}^{\mathrm{tick}}$ is $N \times N$, but the $N$-size intermediate values are stored in the separated memory (see APPENDIX B for details). When a market feed (informing the change of $ask$ or $bid$ of a stock) arrives, the SBM core updates the ($J^{\mathrm{tick}}$, $h^{\mathrm{tick}}$) intermediate data [the size is $\mathcal{O}(N)$] with keeping the ($J^{\mathrm{day}}$, $h^{\mathrm{day}}$) data [the size is $\mathcal{O}(N^2)$].

\section{System}\label{Sec_System}\label{sec_system}
The real-time stock trading system is a hybrid FPGA/CPU system, featuring an event-driven SBM module that starts processing the discrete optimization involved in the proposed strategy when detecting the changes in $ask$ or $bid$ of tradable stocks. The system-wide latency from the market feed arrival to order packet issuance is shortened by co-integrating, in the FPGA, the SBM module together with other system components including communication interfaces. The processing units and memory subsystems in the basic SBM circuit design~\cite{FPL19} have been customized (modified) for the proposed strategy to further improve the system latency.

\subsection{Architecture}\label{subsec_Archi}
\begin{figure}[t]
\centering
\includegraphics[width=8.3 cm]{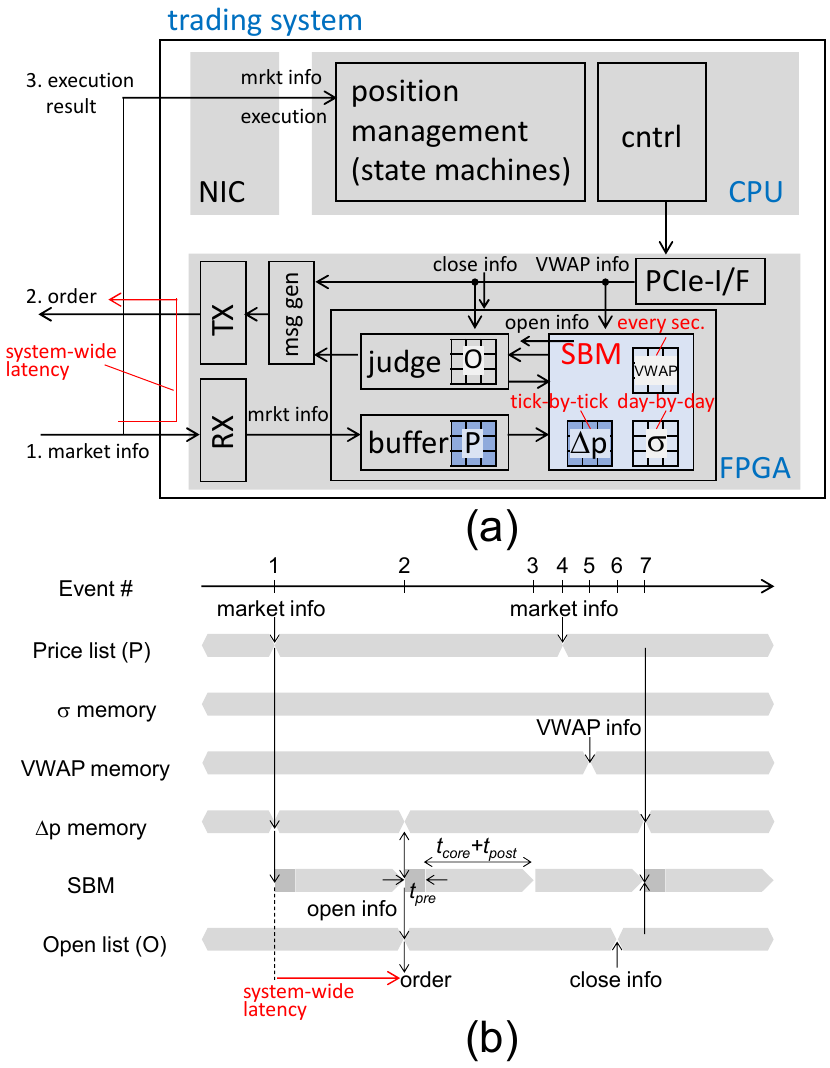}
%\vspace{-0.6cm} 
\caption{System architecture (a hybrid FPGA/CPU system). (a) Block diagram. (b) Timing chart.}
\label{Fig_system}
%\vspace{-0.1cm}
\end{figure}

Figure~\ref{Fig_system} (a) and (b) show the block diagram and timing chart of the hybrid FPGA/CPU system. The FPGA part responds to the changes in the market in a low latency, i.e., it receives the market information, determines the opening of positions based on the NP-hard portfolio optimization by the SBM module, and then issues the order packets. The CPU part controls the whole system and manages the positions using state machines for opened positions (the closing of the positions is determined by the CPU part). The market information (including the changes in $ask$ or $bid$) is received by both the FPGA and CPU parts. The order (buying/selling) packets are issued only from the FPGA part. The execution-result packets informing the results (fill/lapse) of the orders are received by the CPU part. The FPGA and CPU parts are connected with the peripheral component interconnect-express (PCIe) bus.

The system components in the FPGA part are, in the order of data flow, a receiver (RX), a price buffer ($P$) that accommodates the price list of $ask$ or $bid$ for the $N$-stock universe, the SBM module including the three memory units ($\Delta p$, $\sigma$, $VWAP$) which are updated at different timing, a judgment module, a message generator, and a transmitter (TX). Those components are implemented as independent (not synchronized) circuit modules, which are connected with directed streaming data channels with FIFO (first-in-first-out) buffers.

One of the characteristics of the SBM module is that it has three memory units ($\Delta p$, $\sigma$, and $VWAP$) to store data updated at the three timing of tick-by-tick, day-by-day, and every second. The VWAP information, a $\{VWAP_{i}\}$ vector, is updated in the CPU part and informed to the FPGA part at one-second intervals. The SBM module also has a preprocessing submodule ($pre$) to generate the Ising problem described by ($J^{\mathrm{day}}$, $h^{\mathrm{day}}$) and ($J^{\mathrm{tick}}$, $h^{\mathrm{tick}}$) based on the data in the three memory units. The data in the $\Delta p$ memory is changed depending also on the open list ($O$ memory) in the judgment module. The open position is registered in the open list when the opening is decided (before the issuance of the order) and deregistered from there when the closing of the position is confirmed with the message from the CPU part. The $\Delta p$ of the stocks listed in the $O$ memory is set to zero as the duplicate opening is prohibited.

Figure~\ref{Fig_system} (b) shows the timing chart for the operation of the SBM module when representative events (Events $1$ to $7$) happen. When no event happens for a certain time, the SBM module is idling (polling to the FIFO buffers from the price buffer, judgment, and PCIe I/F modules). When a market feed arrives (Event $1$), the SBM module immediately starts the preprocessing (updating of the $\Delta p$ memory and the $J^{\mathrm{tick}}$/$h^{\mathrm{tick}}$ memory) and then the main processing (the discrete optimization). After that, the SBM module evaluates the solution output from the core circuit in terms of the constraint violation and the objective function and then informs the open candidates to the judgment module if the evaluation passes (postprocessing, $post$). After checking the open candidates, the judgment module finally determines the open positions, registers them in the $O$ memory, and concurrently issues order packets via the message generator (Event $2$).

Regardless of whether or not order packets are issued, the SBM module repeats the main processing for a predetermined number of times with different initial states generated by an internal random number generator~\cite{marsaglia03}. As the simulated bifurcation is a heuristic algorithm, the SBM module may find a better solution or another solution enough for the opening. When repeating the main processing, the SBM module also repeats the preprocessing if the order packets have been issued at the last run (in the case of Event $2$) since the $\Delta p$ memory ($O$ memory) has been changed, but skips the preprocessing otherwise (Event $3$). When the Ising problem changes during the main processing because of the arrivals of the new market feed (Event $4$), the VWAP updating information (Even $5$), or the close confirmation information (Event $6$), the information is incorporated at the beginning of the next execution of the SBM module (Event $7$).

\subsection{Customized SBM core circuit}\label{subsec_Core}
To reduce the data size transferred to the SBM core from $\mathcal{O}(N^2)$ to $\mathcal{O}(N)$ (for improving the system latency) when the market situation (or the internal state) changes, we prepare additional computation and memory units to the basic SBM design. Instead of combining the tick-by-tick ($N$-size) and day-by-day  ($N\times N$-size) change information as the coefficients describing the Ising problem, we store that information in separated memory units, separately calculate the updating components of internal variables (corrections of momenta) coming from those two components and then combine the updating components (the number of the components is $N$).

\begin{figure}[t]
\centering
\includegraphics[width=8.3 cm]{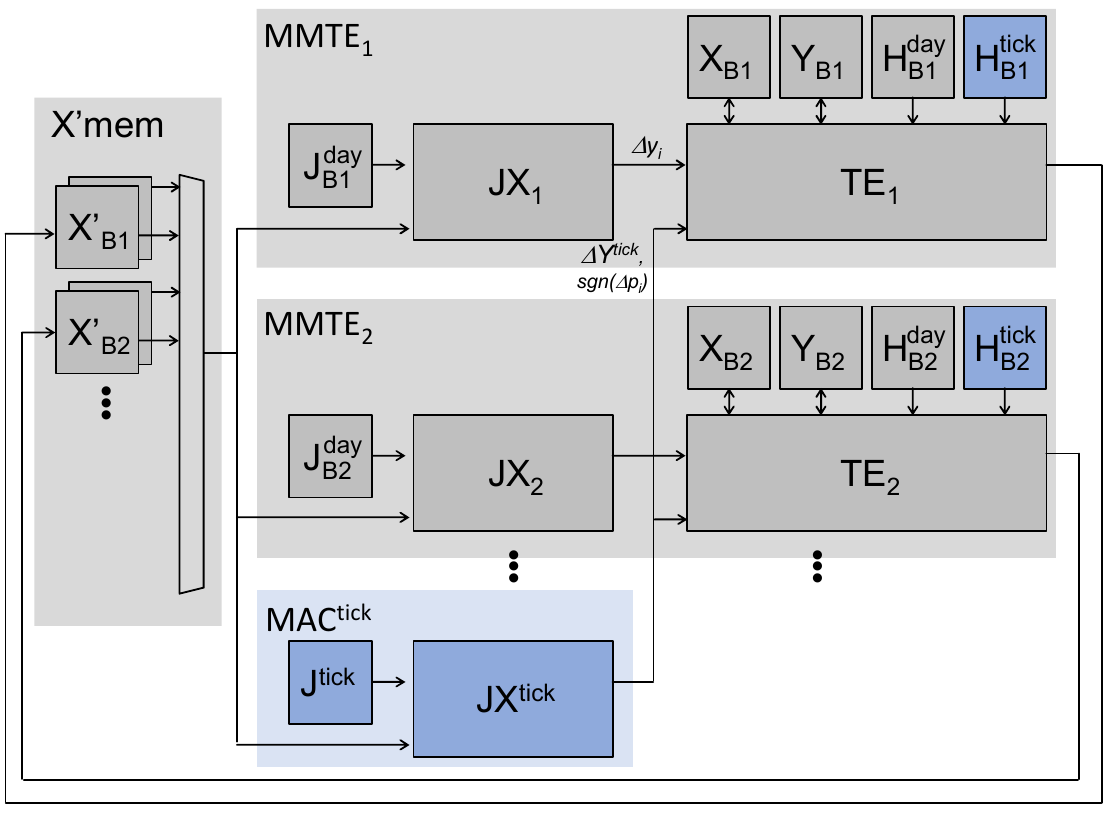}
%\vspace{-0.6cm} 
\caption{Circuit architecture of the SBM core. To shorten the data input time, additional circuit units (blue highlighted) for the computation depending on the tick-by-tick data are introduced to the basic SBM design~\cite{FPL19}.}
\label{Fig_circuit}
%\vspace{-0.1cm}
\end{figure}

Simulated bifurcation~\cite{sbm1, sbm2} simulates the time evolution of $N$ nonlinear oscillators according to the Hamiltonian equations of motion, where the nonlinear oscillators correspond to the spin variables and the state of $i$th oscillator is described by the position and momentum ($x_{i}$, $y_{i}$). The SB time evolution step consists of calculating the correction of momenta $\{\Delta y_{i}\}$ based on the many-body interaction [computationally corresponding to the matrix-vector multiplication (MM) of the $J$ coupling matrix and $\{ x_{i}\}$ position vector] and calculating the updated (time-evolved) state variables, $\{ x_{i}^{k+1}\}$ and $\{ y_{i}^{k+1}\}$, from the $\{\Delta y_{i}\}$, $\{ h_{i}\}$, and the current state variables, $\{ x_{i}^{k}\}$ and $\{ y_{i}^{k}\}$. The additional circuits calculate the correction of the momentum depending only on the tick-by-tick $J$ and $h$ components ($\Delta y_{i}^{\mathrm{tick}}$).

Figure~\ref{Fig_circuit} shows the block diagram of the SB core circuit, where the additional circuit units for the computation depending only on the tick-by-tick change problem components ($J^{\mathrm{tick}}$, $h^{\mathrm{tick}}$) are blue highlighted and the remaining units are architecturally the same as in the basic SBM design~\cite{FPL19}. In the basic design, the main computation components are $\mathrm{JX}$ units corresponding to the multiply-accumulate (MAC) operations of $\sum_{j=1}^{N}J_{ij}x_{j}$ and $\mathrm{TE}$ units corresponding to the time-evolution operation, which are combined to be $\mathrm{MMTE}$ units (each responsible for updating a subgroup of coupled oscillators). The $\mathrm{MMTE}$ units are organized with the global $\mathrm{X^{\prime}_{mem}}$ memory unit to make a circulative structure as a whole corresponding to the iteration of the SB time-evolution steps.

The $J^{\mathrm{tick}}$ and $h^{\mathrm{tick}}$ data are stored in the  $J^{\mathrm{tick}}$ and $H^{\mathrm{tick}}$ memory units, which are separated from the  $J^{\mathrm{day}}$ and $H^{\mathrm{day}}$ memory units storing the day-by-day change problem components ($J^{\mathrm{day}}$, $h^{\mathrm{day}}$). The $JX^{\mathrm{tick}}$ module calculates an intermediate value $\Delta Y^{\mathrm{tick}}$ common for all the oscillators and supply the $\Delta Y^{\mathrm{tick}}$ and $sgn(\Delta p_{i})$ data to the $\mathrm{TE}$ units. The $\mathrm{TE}$ unit calculates the $\Delta y_{i}^{\mathrm{tick}}$ for each oscillator based on the $\Delta Y^{\mathrm{tick}}$,  $sgn(\Delta p_{i})$, $x_i$ and $h_i$ and updates the state of $i$th oscillator with the $\Delta y_{i}^{\mathrm{tick}}$. See APPENDIX B for details.

\begin{figure}[t]
\centering
\includegraphics[width=8.3 cm]{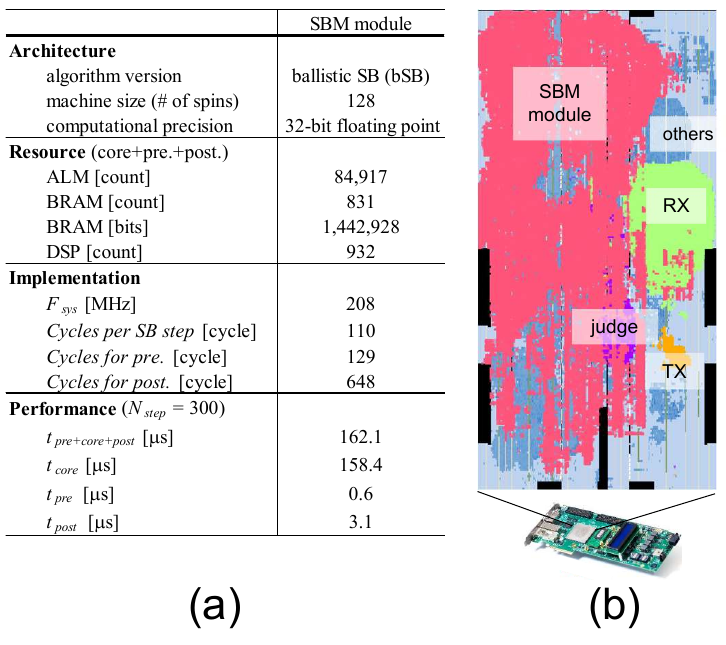}
%\vspace{-0.6cm} 
\caption{System implementation. (a) Architecture and implementation details of the SBM module. (b) Placement of system modules in the FPGA.}
\label{Fig_implementation}
%\vspace{-0.1cm}
\end{figure}

\subsection{Implementation}
We implemented the system described in Sec.~\ref{subsec_Archi} with a CPU server with a network interface card (NIC) and an FPGA board having another network interface (see APPENDIX C for details).

Figure~\ref{Fig_implementation} (a) shows the architecture and implementation results of the SBM module for 128-stock universes ($N$=128). Among three variants of simulated bifurcation (adiabatic, ballistic, and discrete SBs)~\cite{sbm2}, ballistic SB is adopted in this work, with the SB parameters of $N_{\mathrm{step}}$=300 and $dt$=0.02. The machine size (the number of spins) is 128 spins with all-to-all connectively, and the computation precision is 32-bit floating point. Figure~\ref{Fig_implementation} (b) shows the result of the placement of system modules in the FPGA. The SBM module is dominant, and the circuit resources used are listed in Fig.~\ref{Fig_implementation} (a). The system clock frequency determined as a result of circuit synthesis, placement, and routing is 208 MHz. The clock cycles of the SB main (core) processing, preprocessing, and postprocessing are 33,000 per run (110 per SB step), 129, and 648, respectively. The computation time (the module latency) per run ($t_{pre}+t_{core}+t_{post}$) is 162.1 $\mu$s, where the SBM core processing is dominant ($t_{core}$=158.4 $\mu$s). The system-wide latency from the market feed arrival to order packet issuance depicted in Fig.~\ref{Fig_system}(b) as a red arrow is 164 $\mu$s (including the latencies of the RX, price buffer, judgment, SBM, message generator, and TX modules).

\section{Experiment}\label{sec_expt}
The trading system described in Sec.~\ref{sec_system} was installed at the JPX Co-location area of the TSE and operated through real-time trading to examine whether the strategy based on the NP-hard combinatorial optimization proposed in Sec.~\ref{sec_strategy} is executable. The trading results are compared with a backcast simulation of the strategy assuming the orders issued are necessarily filled.

\begin{figure}[t]
\centering
\includegraphics[width=8.3 cm]{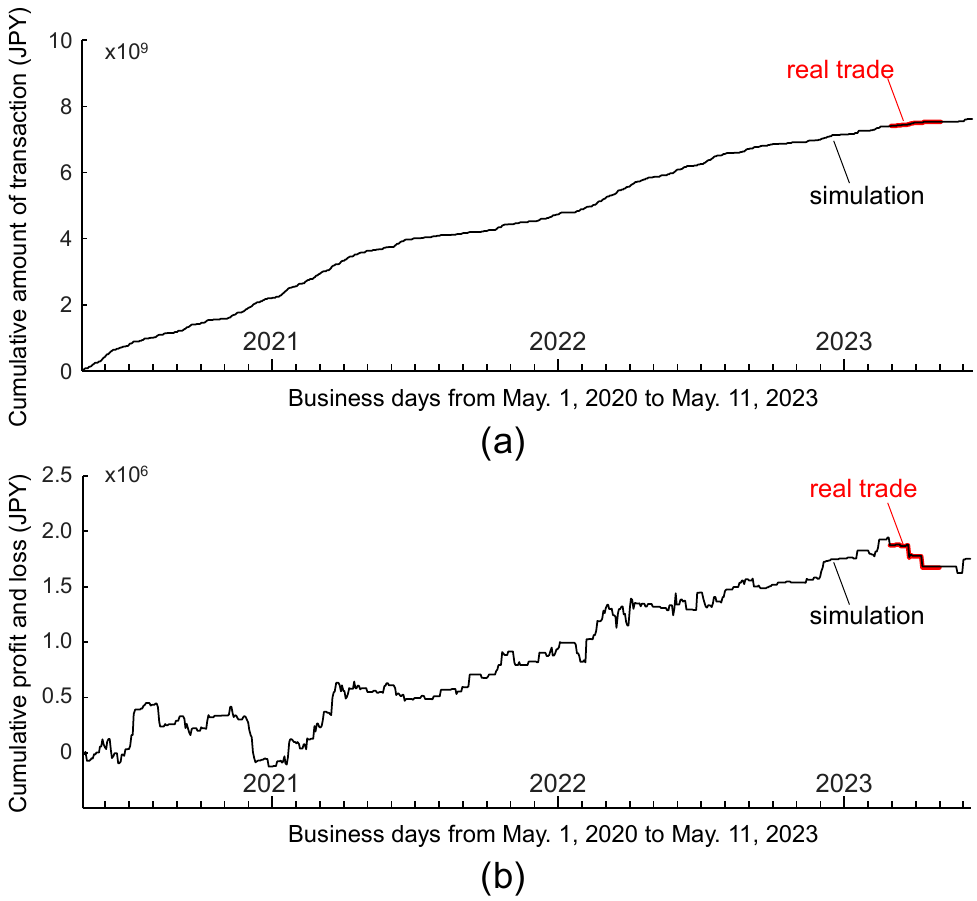}
%\vspace{-0.6cm} 
\caption{Performance of the strategy. (a) Cumulative amount of transactions in JPY and (b) cumulative profit and loss in JPY. Simulation data is from May. 1, 2020, to May. 11, 2023 (737 business days). Real trade data is from Feb. 1, 2023, to Mar. 31, 2023 (42 business days), adjusted with the simulation at the first day.}
\label{Fig_performance}
%\vspace{-0.1cm}
\end{figure}

The proposed strategy determines the opening of positions based on an instantaneous market situation (a price list of $ask$ and $bid$ for the $N$-stock universe). Because of the latency of a system that executes the strategy and the activities of other trading entities, the orders issued are not necessarily filled at the $ask$/$bid$ prices used for the decision-making. We developed a simulator that processes the historical market feeds provided by the TSE and emulates the internal state of the trading system. The simulator assumes that the orders issued are necessarily filled at the intended prices.

Figures \ref{Fig_performance} (a) and (b) show the cumulative values of the amounts of transactions per day and the profit and loss (including $ask$-$bid$ spread costs and commission) per day for real-time trading (red line) and backcast simulation (black line) with fixed strategic parameters of $N$=128, $N_{s}$=4, $P_\mathrm{max}$=4, and $A_\mathrm{trans}$=4 million Japanese yen (JPY). The 128 stocks were selected from the Nikkei 225 or TOPIX 100 constituents in terms of high liquidity. The system is allowed to take positions in the afternoon session of the TSE. The simulation data is from May. 1, 2020, to May. 11, 2023. The real trade data is from Feb. 1, 2023, to Mar. 31, 2023, being adjusted with the simulation at the first day.

The Sharpe ratio of the strategy over the simulation period (approximately 3 years) is 1.23, where the annualized return and risk (the standard deviation of the return) are 3.6 \% and 2.9 \%, respectively, for an investment of 16 million JPY  ($A_\mathrm{trans} \times P_\mathrm{max}$); the strategy proposed can be profitable (a positive annualized return) with a reasonable risk (a low level of annualized risk compared to the annualized return). The cumulative value of the amounts of transactions by the system (118,956,828 JPY) over the experiment (252 hours of real-time trading) is coincident well (+0.01 \%) with the simulation value (118,948,300 JPY), indicating that the strategy proposed is executable with the trading system with a latency of 164 $\mu$s. Note that the slight difference in the transaction amounts comes from the executed prices.

\begin{figure}[t]
\centering
\includegraphics[width=8.3 cm]{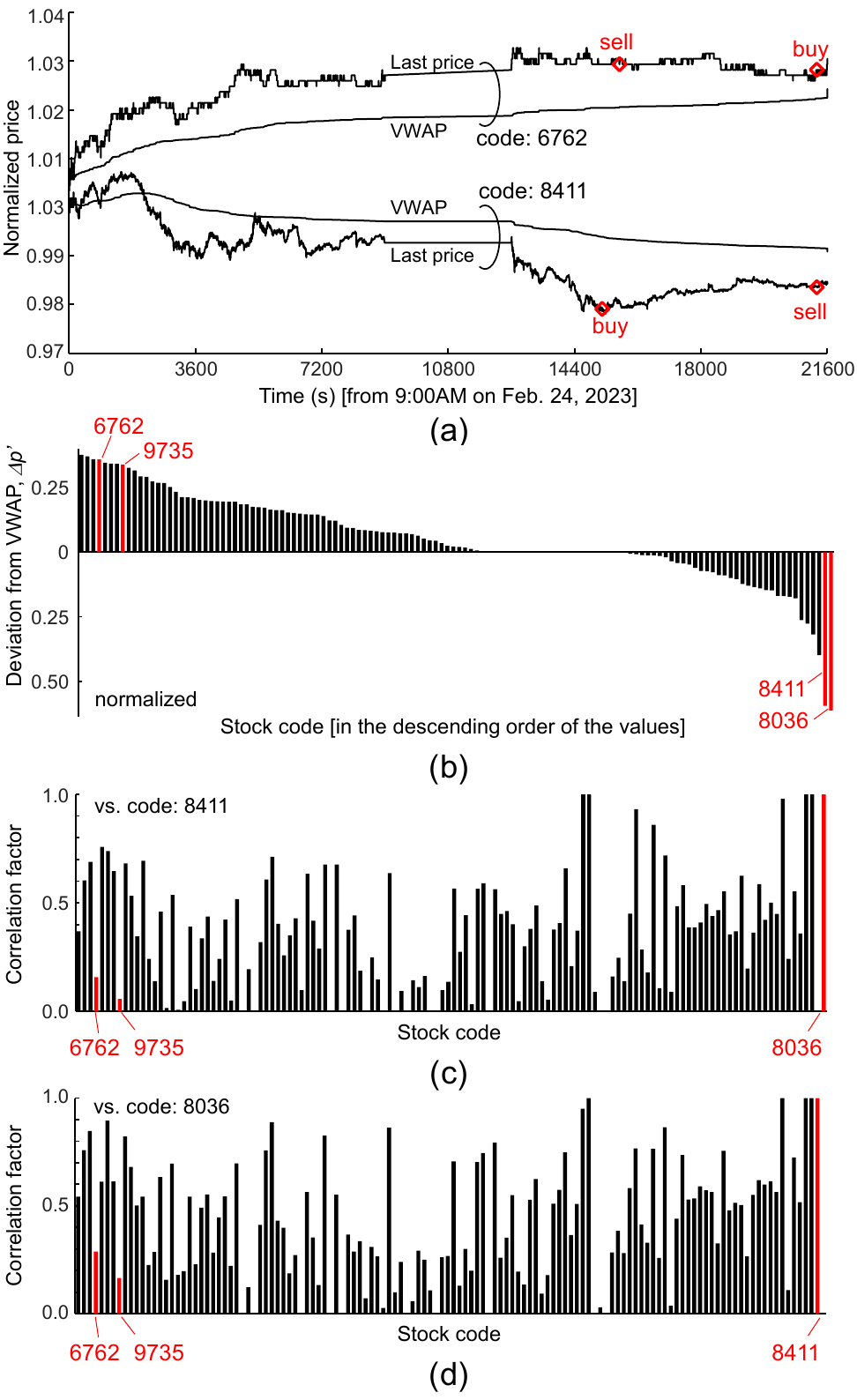}
%\vspace{-0.6cm} 
\caption{A typical transaction by the trading system observed on Feb. 24, 2023. (a) Last price vs. time on the day, where the execution times of transactions are indicated by red markers. The open decision at the time of 15,186 was made based on (b) the deviations from VWAPs of stocks, (c) the correlation factors of stocks vs. code 8411, and (d) the correlation factors of stocks vs. code 8036.}
\label{Fig_typical}
%\vspace{-0.1cm}
\end{figure}

Figure \ref{Fig_typical} shows a typical transaction by the trading system observed on Feb. 24, 2023. On that day, the number of the market feeds informing the changes of $ask$/$bid$ of stocks in the $N(=128)$-stock universe was 5,565,723, which arrived at intervals of 3.6 ms on average. The system decided the opening of the positions at 1:13 PM in JST (15,186 seconds after 9:00 AM) based on the selection of codes 8411, 6762, 8036, and 9735 by the SBM module, leading to the profitable closing of the positions before the end of the day [Fig. \ref{Fig_typical} (a)]. The selection of the four stocks ($N_s=4$) by the SBM module was based on the deviations of the stock prices from the VWAPs ($\Delta p_{i}$) and the correlation factors ($\sigma_{i,j}$) shown in Figs. \ref{Fig_typical} (b), (c), and (d). Codes 8411 and 8036 were selected mainly because of the instantaneous expected returns (the maximum and second-maximum ones at that moment) as the candidates for long positions. From the candidates for short positions balancing to codes 8411 and 8036, codes 6762 and 9735 were chosen based on not only the relatively high expected returns but also relatively low correlation factors against both codes 8411 and 8036. The solution of the SBM module satisfies the constraints of the discrete optimization; $\sum_{i}^{N}b_{i}=N_s$ and $\sum_{i}^{N}sgn(\Delta p_{i})b_{i}=0$ in the representation using bit variables $b_i$.

\section{Conclusion}
We proposed a strategy based on selections of potentially profitable, uncorrelated, and balanced stocks by NP-hard, quadratic and discrete optimization and have demonstrated with the real-time transaction records in the TSE that the strategy is executable in terms of response latency with the automated trading system using the SB-based embeddable Ising machine for the selection problem.

The cost function of $N_{\mathrm{s}}$-stock selection problem is designed to involve maximizing instantaneous expected returns defined as deviations from volume-weighted average prices (VWAPs), minimizing the summation of statistical correlation factors (for correlation diversification), and penalty functions for $N_{\mathrm{s}}$-stock selection and delta-neutral positions. The selection problem is formulated in the form of the Ising problem and then the data describing the problem is separated into two components that change tick-by-tick ($N$-size) or day-by-day ($N\times N$-size). By customizing the SBM core circuit to have two sets of memory and computation modules respectively for the tick-by-tick and day-by-day change problem-components, we reduced the data size transferred to the SBM core from $\mathcal{O}(N^2)$ to $\mathcal{O}(N)$ when the market situation changes and improved the system latency. This is a technique to improve the system latency when the problem components change at different timing and is applicable to the SB algorithm and other algorithms based on Hamiltonian equations of motion.

The automated trading system is a hybrid FPGA/CPU system, featuring an event-driven SBM module in the FPGA part. The FPGA part (hardware processing) decides the opening of a group of long/short positions using the SBM and then issues the corresponding orders, while the CPU part (software processing) manages the positions (including the decision of closing positions). The system-wide latency from the market feed arrival to the order packet issuance is 164 $\mu$s for a $128$-stock universe.

The trading system was installed at the JPX Co-location area of the TSE and operated for a real-time trading period of 42 business days or 252 hours. The real-time transaction records were compared with a backcast simulation of the strategy assuming the orders issued are necessarily filled at the intended prices. Based on the good agreement in the cumulative transaction amounts and detailed comparison analysis of transactions between the experiment and simulation, we have concluded that the response latency of the system with the SB-based Ising machine is sufficiently low to execute the trading strategy based on the NP-hard discrete portfolio optimization.

Automated trading systems with embedded Ising machines would be applicable to the strategies based on various discrete portfolio optimizations characterized by different definitions of expected returns and correlations [diagonal and non-diagonal terms in Eq. \eqref{Eq_qubo_cost}] and other trading strategies that rely on high-speed discrete optimization.

\section*{Appendices}
\subsection*{A. QUBO \& Ising representations}
The QUBO formulation ($b_{i}\in{\{0,1\}}$),
\begin{equation}
H_{\mathrm{QUBO}}=\sum_{i}^{N}\sum_{j}^{N}Q_{i,j}b_{i}b_{j},
\end{equation}
is represented also in the Ising formulation ($s_{i}\in{\{-1,1\}}$) as follows.
\begin{equation}
H_{\mathrm{Ising}}=-\frac{1}{2}\sum_{i}^{N}\sum_{j}^{N}J_{i,j}s_{i}s_{j}+\sum_{i}^{N}h_{i}s_{i},
\end{equation}
where 
\begin{equation}
s_{i}=2b_{i}-1,
\end{equation}
\begin{equation}
J_{i,j}=
	\begin{cases}
		-\frac{Q_{i,j}}{2} & (\text{if}\;i\neq j),\\
		0 & (\text{if}\;i= j),
	\end{cases}
\end{equation}
\begin{equation}
h_{i}=\sum_{j}^{N}\frac{Q_{i,j}}{2}.
\end{equation}

\subsection*{B. Additional computation units}
The correction ($\Delta y_{i}^{\mathrm{tick}}$) of the momentum per SB time-evolution step for $i$th oscillator depending on the tick-by-tick $J$ and $h$ components is expressed by
\begin{equation}
\Delta y_{i}^{\mathrm{tick}}=\frac{c_{3}}{2}\left( \Delta Y^{\mathrm{tick}}-x_{i}\right) sgn(\Delta p_{i})-h_{i}^{\mathrm{tick}},
\end{equation}
where
\begin{equation}
\Delta Y^{\mathrm{tick}}=\sum_{i}^{N} sgn(\Delta p_{i}) x_{i},
\end{equation}
\begin{equation}
h_{i}^{\mathrm{tick}}=-c_1 \frac{\left| \Delta p_{i} \right |}{2} + c_3 \left( \sum_{j}^{N} \frac{sgn(\Delta p_{j})}{2}\right) sgn(\Delta p_{i}).
\end{equation}

The $JX^{\mathrm{tick}}$ in the $MAC^{\mathrm{tick}}$ module (Fig.~\ref{Fig_circuit}) is provided with the $x_{i}$ and $sgn(\Delta p_{i})$ data from the global $\mathrm{X^{\prime}_{mem}}$ memory and the $J^{\mathrm{tick}}$ memory and calculates the $\Delta Y^{\mathrm{tick}}$ in a spatially parallel manner using multiple MAC processing elements. The time evolution (TE) module receives the $\Delta Y^{\mathrm{tick}}$ and $sgn(\Delta p_{i})$ data from the $MAC^{\mathrm{tick}}$ module and also receives the $h_{i}^{\mathrm{tick}}$ data from the $H^{\mathrm{tick}}$ memory, and then updates the momentum of each oscillator by respective correction of $\Delta y_{i}^{\mathrm{tick}}$ in a temporal parallel manner (pipelining).

\subsection*{C. Implementation details}
An FPGA board and a high-speed network interface card (NIC) are mounted on a host server with dual CPUs (Intel Xeon Silver 4215R) and DDR-DRAM modules (384 GB). The FPGA (Intel Arria 10 GX 1150 FPGA) on the board has 427,200 adaptive logic modules (ALMs) including 854,400 adaptive look-up-tables (ALUTs, 5-input LUT equivalent) and 1,708,800 flip-flop registers, 2,713 20Kbit-size RAM blocks (BRAMs), and 1,518 digital signal processor blocks (DSPs). The system components in the FPGA described in Section~\ref{Sec_System} were coded in a high-level synthesis (HLS) language (Intel FPGA SDK for OpenCL, ver.~18.1). The FPGA interfaces including a PCIe IP (PCIe Gen3$\times$8), a 10~Gbps Ethernet PHY IP and communication IPs (RX, TX) were written in Verilog HDL and incorporated in the board support package (BSP).

\subsection*{Acknowledgment}
The experiment in the Tokyo Stock Exchange was conducted under a joint project between Toshiba Corporation and Dharma Capital. K.K. The authors thank  Ryosuke Iio and Kohei Shimane for fruitful discussions and technical support.

\subsection*{Conflicts of Interest}
K.T., R.H., and M.Y. are included in inventors on two U.S. patent applications related to this work filed by the Toshiba Corporation (no. 17/249353, filed 20 February 2020; no. 17/565206, filed 29 December 2021). The authors declare that they have no other competing interests.

\clearpage
\begin{figure}[h]
\vspace{0.5cm}
\noindent\includegraphics[width=1in,height=1.25in,clip,keepaspectratio]{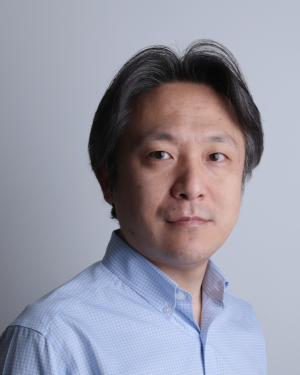}\\
{\small Kosuke Tatsumura received the B.E., M.E., and Ph.D. degrees in Electronics, Information and Communications Engineering from Waseda University, Japan, in 2000, 2001, and 2004, respectively. After working as a postdoctoral fellow at Waseda University, he joined Toshiba Corporation in 2006. He is a chief research scientist, leading a research team and several projects toward realizing innovative industrial systems based on cutting-edge computing technology. He was a member of the Emerging Research Devices (ERD) committee in the International Technology Roadmap for Semiconductors (ITRS) from 2013 to 2015. He has been a lecturer at Waseda University since 2013. He was a visiting researcher at the University of Toronto from 2015 to 2016. He received the Best Paper Award at IEEE Int. Conf. on Field-Programmable Technology (FTP) in 2016. His research interests include domain-specific computing, quantum/quantum-inspired computing, and their applications.}
\vspace{-0.5cm} 
\end{figure}

\begin{figure}[h]
\vspace{0.5cm}
\noindent\includegraphics[width=1in,height=1.25in,clip,keepaspectratio]{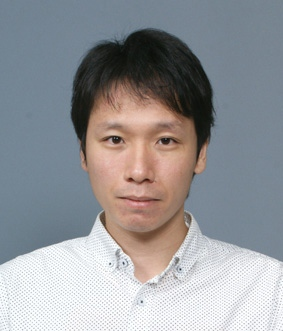}\\
{\small Ryo Hidaka received the B.E. and M.E. degrees in Systems Design and Informatics from Kyushu Institute of Technology, Japan, in 2006 and 2008, respectively. He joined Toshiba Corporation in 2008. He was engaged in the development of main processors (2D-to-3D conversion and local dimming) for digital televisions, an image recognition processor called Visconti\textsuperscript{TM}, and host controllers for flash-memory cards. His current research interests include domain-specific computing, high-level synthesis design methodology, and proof-of-concept study with FPGA devices.}
\vspace{-0.5cm} 
\end{figure}

\begin{figure}[h]
\vspace{0.5cm}
\noindent\includegraphics[width=1in,height=1.25in,clip,keepaspectratio]{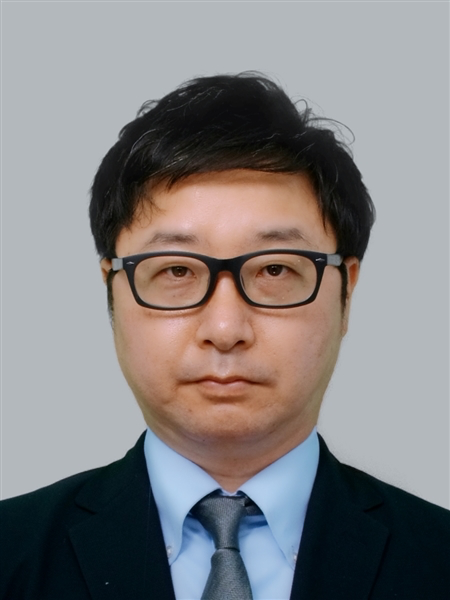}\\
{\small Jun Nakayama received the degree of Bachelor of Arts in Economic and Social studies from the University of Manchester, the U.K., in 2008. He received the degree of Master of Business Administration (MBA) in Finance from Hitotsubashi University, Japan, in 2017. He was a portfolio manager in Nomura Asset Management Co., Ltd. from 2008 to 2020 and was engaged in the development and management of quant-based funds. He joined Toshiba Corporation in 2020. He is also a Ph.D. candidate in the Financial Strategy Program, Hitotsubashi University Business School. His research interests include quantitative investment strategies, quantum-inspired computing technology, and trading strategies with advanced technologies.}
\vspace{-0.5cm} 
\end{figure}

\begin{figure}[h]
\vspace{0.5cm}
\noindent\includegraphics[width=1in,height=1.25in,clip,keepaspectratio]{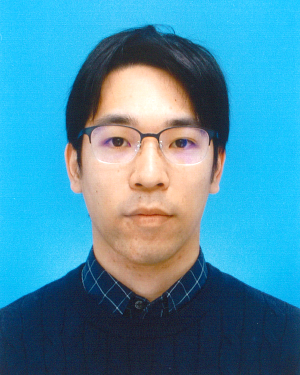}\\
{\small Tomoya Kashimata received the B.E., and M.E. degrees in computer science and engineering from Waseda University, Japan, in 2018 and 2020, respectively. He joined Corporate Research and Development Center, Toshiba Corporation, Japan, in 2020. His research interests include computer architecture, reconfigurable architecture, and processor in memory.}
\vspace{-0.5cm} 
\end{figure}

\begin{figure}[h]
\vspace{0.5cm}
\noindent\includegraphics[width=1in,height=1.25in,clip,keepaspectratio]{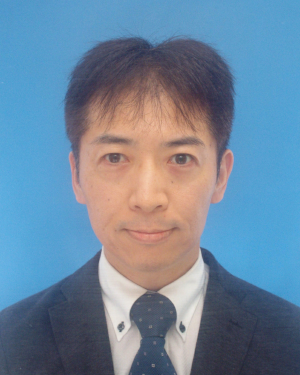}\\
{\small Masaya Yamasaki received the B.E. and M.E. degrees in Computer Science and Communication Engineering from Kyushu University, Japan, in 1997 and 1999, respectively. He joined Toshiba Corporation in 1999. He was engaged in the development of image processing engines (interframe interpolation technology) for digital televisions (including ones with Cell Broadband Engines) and FPGA-based coprocessors for multi-channel video recording, three-dimensional display, and industrial systems. His research interests include domain-specific computing, high-level synthesis design space exploration, and proof-of-concept study with FPGA devices.}
\vspace{-0.5cm} 
\end{figure}


\begin{thebibliography}{00}
{\small 
%discrete portfolio optimizations are NP-hard
\bibitem{bienstock96} D. Bienstock, ``Computational study of a family of mixed-integer quadratic programming problems,'' \emph{Mathematical programming} \textbf{74}, pp. 121--140, 1996. [Online]. Available: https://doi.org/10.1007/BF02592208

\bibitem{mansini99} R. Mansini, M. G. Speranza, ``Heuristic algorithms for the portfolio selection problem with minimum transaction lots,'' \emph{European Journal of Operational Research} \textbf{114}, pp. 219--233, 1999. [Online]. Available: https://doi.org/10.1016/S0377-2217(98)00252-5

%NP-hard optimizations
%standard
\bibitem{markowitz52} H. Markowitz, ``Portfolio selection,'' \emph{The Journal of Finace} \textbf{7}, pp. 77--91, 1952. [Online]. Available: https://doi.org/10.2307/2975974

\bibitem{venturelli19} D. Venturelli, A. Kondratyev, ``Reverse quantum annealing approach to portfolio optimization problems,'' \emph{Quantum Machine Intelligence} \textbf{1}, pp. 17--30, 2019. [Online]. Available: https://doi.org/10.1007/s42484-019-00001-w

\bibitem{lang22} J. Lang, S. Zielinski, S. Feld, ``Strategic Portfolio Optimization Using Simulated, Digital, and Quantum Annealing,'' \emph{Applied Sciences} \textbf{12}, 12288, 2022. [Online]. Available: https://doi.org/10.3390/app122312288

%optimal trading trajectory problem
\bibitem{rosenberg15} G. Rosenberg, P. Haghnegahdar, P. Goddard, P. Carr, K. Wu, M. L. De Prado, ``Solving the optimal trading trajectory problem using a quantum annealer,'' \emph{Proc. of Workshop on High Performance Computational Finance} (WHPCF), pp. 1--7, 2015. [Online]. Available: https://doi.org/10.1145/2830556.2830563

\bibitem{steinhauer20} K. Steinhauer, T. Fukadai, S. Yoshida, ``Solving the Optimal Trading Trajectory Problem Using Simulated Bifurcation,'' \emph{arXiv preprint} arXiv:2009.08412, 2020. [Online]. Available: https://doi.org/10.48550/arXiv.2009.08412

\bibitem{mugel22} S. Mugel, C. Kuchkovsky, E. Sanchez, S. Fernandez-Lorenzo, J. Luis-Hita, E. Lizaso, R. Orus, ``Dynamic portfolio optimization with real datasets using quantum processors and quantum-inspired tensor networks,'' \emph{Physical Review Research}  \textbf{4}, 013006, 2022. [Online]. Available: https://doi.org/10.1103/PhysRevResearch.4.013006

%diversified portfolio
\bibitem{butenko03} S. Butenko,, ``Maximum independent set and related problems, with applications,'' \emph{Ph.D. dissertation, the Industrial and Systems Engineering Department, University of Florida}, 2003. [Online]. Available: https://ufdcimages.uflib.ufl.edu/UF/E0/00/10/11/\\00001/butenko\_s.pdf

\bibitem{boginski04} V. Boginski,  S. Butenko, P. M. Pardalos,  ``Network-based Techniques in the Analysis of the Stock Market,'' in \emph{Supply Chain and Finance}, eds. P. M. Pardalos, A. Migdalas, G. Baourakis, World Scientific, pp. 1--14, 2004. [Online]. Available: https://doi.org/10.1142/9789812562586\_0001

\bibitem{marzec16} M. Marzec, ``Portfolio optimization: Applications in quantum computing,'' in \emph{Handbook of High-Frequency Trading and Modeling in Finance} eds. I. Florescu, M. C. Mariani, H. E. Stanley, F. G. Viens, Wiley Online Library, pp. 73--106, 2016. [Online]. Available: https://doi.org/10.1002/9781118593486.ch4

\bibitem{sakurai21} Y. Sakurai, Y. Yuki, R. Katsuki, T. Yazane, F. Ishizaki, ``Correlation Diversified Passive Portfolio Strategy Based on Permutation of Assets,'' \emph{Journal of Investment Strategies} \textbf{10}, pp. 1--22, 2021. [Online]. Available: http://doi.org/10.21314/JOIS.2021.010

%Ising machine
%SBM
\bibitem{sbm1} H. Goto, K. Tatsumura, A. R. Dixon, ``Combinatorial optimization by simulating adiabatic bifurcations in nonlinear Hamiltonian systems,'' \emph{Science Advances} \textbf{5}, eaav2372, 2019. [Online]. Available: https://doi.org/10.1126/sciadv.aav2372

\bibitem{FPL19} K. Tatsumura, A. R. Dixon, H. Goto, ``FPGA-Based Simulated Bifurcation Machine,'' \emph{Proc. of IEEE International Conference on Field Programmable Logic and Applications} (FPL), pp. 59--66, 2019. [Online]. Available: https://doi.org/10.1109/FPL.2019.00019

\bibitem{sbm2} H. Goto, K. Endo, M. Suzuki, Y. Sakai, T. Kanao, Y. Hamakawa, R. Hidaka, M. Yamasaki, K. Tatsumura, ``High-performance combinatorial optimization based on classical mechanics,'' \emph{Science Advances} \textbf{7}, eabe7953, 2021. [Online]. Available: https://doi.org//10.1126/sciadv.abe7953

\bibitem{NatEle} K. Tatsumura, M. Yamasaki, H. Goto, ``Scaling out Ising machines using a multi-chip architecture for simulated bifurcation,'' \emph{Nature Electronics} \textbf{4}, pp. 208--217, 2021. [Online]. Available: https://doi.org/10.1038/s41928-021-00546-4

\bibitem{kanao23} T. Kanao, H. Goto, ``Simulated bifurcation for higher-order cost functions,'' \emph{Applied Physics Express} \textbf{16}, 014501, 2023. [Online]. Available: https://doi.org/10.35848/1882-0786/acaba9

%D-wave
\bibitem{johnson11}
M. W. Johnson, M. H. S. Amin, S. Gildert, T. Lanting, F. Hamze, N. Dickson, R. Harris, A. J. Berkley, J. Johansson, P. Bunyk, E. M. Chapple, C. Enderud, J. P. Hilton, K. Karimi, E. Ladizinsky, N. Ladizinsky, T. Oh, I. Perminov, C. Rich, M. C. Thom, E. Tolkacheva, C. J. S. Truncik, S. Uchaikin, J. Wang, B. Wilson, G. Rose, ``Quantum annealing with manufactured spins,'' \emph{Nature} \textbf{473}, pp. 194--198 (2011). [Online]. Available: https://doi.org/10.1038/nature10012

\bibitem{king23}
A. D. King, J. Raymond, T. Lanting, R. Harris, A. Zucca, F. Altomare, A. J. Berkley, K. Boothby, S. Ejtemaee, C. Enderud, E. Hoskinson, S. Huang, E. Ladizinsky, A. J. R. MacDonald, G. Marsden, R. Molavi, T. Oh, G. Poulin-Lamarre, M. Reis, C. Rich, Y. Sato, N. Tsai, M. Volkmann, J. D. Whittaker, J. Yao, A. W. Sandvik, M. H. Amin, ``Quantum critical dynamics in a 5,000-qubit programmable spin glass,'' \emph{Nature} \textbf{617}, pp. 61–-66 (2023). [Online]. Available: https://doi.org/10.1038/s41586-023-05867-2

%hybrid optics/electronics systems
\bibitem{honjo21}
T. Honjo, T. Sonobe, K. Inaba, T. Inagaki, T. Ikuta, Y. Yamada, T. Kazama, K. Enbutsu, T. Umeki, R. Kasahara, K. Kawarabayashi, H. Takesue, ``100,000-spin coherent ising machine,'' \emph{Science Advances} \textbf{7}, eabh095 (2021). [Online]. Available: https://doi.org/10.1126/sciadv.abh0952

\bibitem{pierangeli19}
D. Pierangeli, G. Marcucci, C. Conti, ``Large-Scale Photonic Ising Machine by Spatial Light Modulation,'' \emph{ Physical Review Letters} \textbf{122}, 213902 (2019). [Online]. Available: https://doi.org/10.1103/PhysRevLett.122.213902

%memristor Hopfield neural networks
\bibitem{cai20} F. Cai, S. Kumar, T. V. Vaerenbergh, X. Sheng, R. Liu, C. Li, Z. Liu, M. Foltin, S. Yu, Q. Xia, J. J. Yang, R. Beausoleil, W. D. Lu, J. P. Strachan, ``Power-efficient combinatorial optimization using intrinsic noise in memristor Hopfield neural networks,'' \emph{Nature Electronics} \textbf{3}, pp. 409--418, 2020. [Online]. Available: https://doi.org/10.1038/s41928-020-0436-6

%p-bit
\bibitem{aadit22} N. A. Aadit, A. Grimaldi, M. Carpentieri, L. Theogarajan, J. M. Martinis, G. Finocchio, K. Camsari, ``Massively parallel probabilistic computing with sparse Ising machines,'' \emph{Nature Electronics} \textbf{5}, pp. 460--468, 2022. [Online]. Available: https://doi.org/10.1038/s41928-022-00774-2

%coupled oscillator circuit
\bibitem{moy22} W. Moy, I. Ahmed, P. Chiu, J. Moy, S. S. Sapatnekar, C. H. Kim, ``A 1,968-node coupled ring oscillator circuit for combinatorial optimization problem solving,'' \emph{Nature Electronics} \textbf{5}, pp. 310--317, 2022. [Online]. Available: https://doi.org/10.1038/s41928-022-00749-3

%analog ASIC
\bibitem{sharma22} A. Sharma, R. Afoakwa, Z. Ignjatovic, M. Huang, ``Increasing Ising machine capacity with multi-chip architectures,'' \emph{Proc. of Annual International Symposium on Computer Architecture} (ISCA), pp. 508--521, 2022. [Online]. Available: https://doi.org/10.1145/3470496.3527414

%digital ASIC
\bibitem{takemoto19}
T. Takemoto, M. Hayashi, C. Yoshimura, M. Yamaoka, ``A 2$\times$30k-Spin Multi-Chip Scalable Annealing Processor Based on a Processing-In-Memory Approach for Solving Large-Scale Combinatorial Optimization Problems,'' \emph{IEEE Journal of Solid-State Circuits} \textbf{55}, pp. 145--156, 2019. [Online]. Available: https://doi.org/10.1109/JSSC.2019.2949230

\bibitem{kawamura23} K. Kawamura, J. Yu, D. Okonogi, S. Jimbo, G. Inoue, A. Hyodo, {\'A}. L. Garc{\'\i}a-Anas, K. Ando, B. H. Fukushima-Kimura, R. Yasudo, T. Van Chu, M. Motomura, ``Amorphica: 4-replica 512 fully connected spin 336MHz metamorphic annealer with programmable optimization strategy and compressed-spin-transfer multi-chip extension,'' {Proc. of  IEEE International Solid-State Circuits Conference} (ISSCC), pp. 42--43, 2023. [Online]. Available: https://doi.org/10.1109/ISSCC42615.2023.10067504

%DAU 2gen
\bibitem{matsubara20} S. Matsubara, M. Takatsu, T. Miyazawa, T. Shibasaki, Y. Watanabe, K. Takemoto, H. Tamura, ``Digital annealer for high-speed solving of combinatorial optimization problems and its applications,'' \emph{Proc. of Asia and South Pacific Design Automation Conference} (ASP-DAC), pp. 667--672, 2020. [Online]. Available: https://doi.org/10.1109/ASP-DAC47756.2020.9045100

%FPGA
\bibitem{waidyasooriya21} H. M. Waidyasooriya, M. Hariyama, ``Highly-parallel FPGA accelerator for simulated quantum annealing,'' \emph{ IEEE Transactions on Emerging Topics in Computing} \textbf{9}, pp. 2019–2029, 2021. [Online]. Available: https://doi.org/10.1109/TETC.2019.2957177

%GPU
\bibitem{okuyama19} T. Okuyama, T. Sonobe, K. Kawarabayashi, M. Yamaoka, ``Binary optimization by momentum annealing,'' \emph{ Physical Review E} \textbf{100}, 012111, 2019. [Online]. Available: https://doi.org/10.1103/PhysRevE.100.012111

%about Ising machines
%Ising is NP-hard
\bibitem{barahona82} F. Barahona, ``On the computational complexity of Ising spin glass models,'' \emph{Journal of Physics A: Mathematical and General} \textbf{15}, pp. 3241–-3253, 1982. [Online]. Available: https://doi.org/10.1088/0305-4470/15/10/028

\bibitem{lucas14} A. Lucas, ``Ising formulations of many NP problems,'' \emph{Frontiers in physics} \textbf{2}, 5, 2014. [Online]. Available: https://doi.org/10.3389/fphy.2014.00005

\bibitem{SA83} S. Kirkpatrick, C. D. Gelatt, M. P. Vecchi, ``Optimization by simulated annealing,'' \emph{Science} \textbf{220}, pp. 671–-680, 1983. [Online]. Available: https://doi.org/10.1126/science.220.4598.671

%trading systems
\bibitem{yoo23} S. Yoo, H. Kim, J. Kim, S. Park, J.-Y. Kim, J. Oh, ``LightTrader: A Standalone High-Frequency Trading System with Deep Learning Inference Accelerators and Proactive Scheduler,'' \emph{IEEE International Symposium on High-Performance Computer Architecture} (HPCA), pp. 1017--1030, 2023. [Online]. Available: https://doi.org/10.1109/HPCA56546.2023.10070930

\bibitem{fil20} M. Fil, L. Kristoufek, ``Pairs trading in cryptocurrency markets,'' \emph{IEEE Access} \textbf{8}, pp. 172644--172651, 2020. [Online]. Available: https://doi.org/10.1109/ACCESS.2020.3024619

\bibitem{huang19} B. Huang, Y. Huan, L. D. Xu, L. Zheng, Z. Zou, ``Automated trading systems statistical and machine learning methods and hardware implementation: a survey,'' \emph{Enterprise Information Systems} \textbf{13}, pp. 132--144, 2019. [Online]. Available: https://doi.org/10.1080/17517575.2018.1493145

\bibitem{denholm15} S. Denholm, H. Inoue, T. Takenaka, T. Becker, W. Luk, ``Network-level FPGA acceleration of low latency market data feed arbitration,'' \emph{IEICE Transactions on Information and Systemss} \textbf{E98-D}, pp. 288--297, 2015. [Online]. Available: https://doi.org/10.1587/transinf.2014RCP0011

\bibitem{leber11} C. Leber, B. Geib, H. Litz, ``High frequency trading acceleration using FPGAs,'' \emph{Proc. of IEEE International Conference on Field Programmable Logic and Applications} (FPL), pp. 317--322, 2011. [Online]. Available: https://doi.org/10.1109/FPL.2011.64

%modern FPGA architecture
\bibitem{betz12} V. Betz, J. Rose, A. Marquardt, ``Architecture and CAD for deep-submicron FPGAs,'' Springer New York, NY, 1999  [Online]. Available: https://doi.org/10.1007/978-1-4615-5145-4

\bibitem{ISCAS20} K. Tatsumura, R. Hidaka, M. Yamasaki, Y. Sakai, H. Goto, ``A Currency Arbitrage Machine based on the Simulated Bifurcation Algorithm for Ultrafast Detection of Optimal Opportunity,'' \emph{Proc. of IEEE International Symposium on Circuits and Systems} (ISCAS), pp. 1--5, 2020. [Online]. Available: https://doi.org/10.1109/ISCAS45731.2020.9181114

%QBM
\bibitem{qbm} H. Goto, ``Bifurcation-based adiabatic quantum computation with a nonlinear oscillator network,'' \emph{Scientific Reports} \textbf{6}, 21686, 2016. [Online]. Available: https://doi.org/10.1038/srep21686

%The importance of automated trading systems
\bibitem{malceniece23} L. Malceniece, K. Malcenieks, T. J. Putni{\c{n}}{\v{s}}, T{\=a}lis, ``High frequency trading and comovement in financial markets,'' \emph{Journal of Financial Economics} \textbf{134}, pp. 381--399, 2019. [Online]. Available: https://doi.org/10.1016/j.jfineco.2018.02.015

\bibitem{brogaard14} J. Brogaard, T. Hendershott, R. Riordan, ``High-Frequency Trading and Price Discovery,'' \emph{The Review of Financial Studies} \textbf{27}, pp. .2267-2306, 2014. [Online]. Available: https://doi.org/10.1093/rfs/hhu032

%Sharpe ratio
\bibitem{sharpe66} W. F. Sharpe,  ``Mutual fund performance,'' \emph{The Journal of Business} \textbf{39}, pp. 119--138, 1966. [Online]. Available: https://www.jstor.org/stable/2351741

%the def of Sharpe ratio
\bibitem{backus93} D. K. Backus, A. W. Gregory, C. I. Telmer, ``Accounting for forward rates in markets for foreign currency,'' \emph{The Journal of Finance} \textbf{48}, pp. 1887--1908, 1993. [Online]. Available: https://doi.org/10.1111/j.1540-6261.1993.tb05132.x

%Pairs Trade
\bibitem{gatev06} E. Gatev, W. N. Goetzmann, K. G. Rouwenhorst, ``Pairs trading: Performance of a relative-value arbitrage rule,'' \emph{The Review of Financial Studies} \textbf{19}, pp. 797--827, 2006. [Online]. Available: https://doi.org/10.1093/rfs/hhj020

%VWAP as the benchmark
\bibitem{berkowitz88} S. A. Berkowitz, D. E. Logue,  E. A. Noser Jr, ``The total cost of transactions on the NYSE,'' \emph{The Journal of Finance} \textbf{43}, pp. 97--112, 1988. [Online]. Available: https://doi.org/10.1111/j.1540-6261.1988.tb02591.x

\bibitem{kakade04} S. M. Kakade, M. Kearns, Y. Mansour, L. E. Ortiz, ``Competitive algorithms for VWAP and limit order trading,'' \emph{Proc. of ACM conference on Electronic commerce}, pp. 189--198, 2004. [Online]. Available: https://doi.org/10.1145/988772.988801

\bibitem{bialkowski08} J. Bia{\l}kowski, S. Darolles, G. Le Fol, ``Improving VWAP strategies: A dynamic volume approach,'' \emph{Journal of Banking \& Finance} \textbf{32}, pp. 1709--1722, 2008. [Online]. Available: https://doi.org/10.1016/j.jbankfin.2007.09.023

%RNG
\bibitem{marsaglia03} G. Marsaglia, ``Xorshift RNGs,'' \emph{Journal of Statistical software} \textbf{8}, pp. 1--6, 2003. [Online]. Available: https://doi.org/10.18637/jss.v008.i14
}
\end{thebibliography}
\end{document}